# Au/Ag bimetallic nanocomposites as a highly sensitive plasmonic material

**Taerin Chung, Charles Soon Hong Hwang, Myeong-Su Ahn, and Ki-Hun Jeong***

Department of Bio and Brain engineering and KAIST Institute for Health Science and Technology, Korea Advanced Institute of Science and Technology (KAIST), 291 Dahak-ro, Yuseong-gu, Daejeon, 34141, Republic of Korea
*kjeong@kaist.ac.kr

## ABSTRACT

We report Au/Ag bimetallic nanocomposites as a highly sensitive plasmonic material. A unit approach via a three-dimensional numerical modeling is introduced to observe collective plasmon resonance in Au/Ag bimetallic nanocomposites as well as Au mono-metallic nanoensembles. Au nanoensembles provide consistently identical plasmon wavelength, independent of inter-unit distance. In analogy with mono-metallic nanoensembles, Au/Ag bimetallic nanocomposites distinctly feature converging dual plasmon resonance peaks to a single plasmon resonance peak, strongly depending on the packing density and the unit size. An effective unit size of bimetallic nanocomposites is below 2.5 nm in a subwavelength structure, which is small enough to feature bimetallic nanocomposites. As a result, the Au/Ag bimetallic nanocomposites clearly show exceptionally high sensitivity and figure-of-merit (approximately 3 fold of conventional plasmon sensitivity and 4.3 fold of conventional plasmon FOM), resulting from coupled Au-Ag quadrupole bimetallic nanounits. This study provides essential rationales for Au/Ag bimetallic nanocomposites serving as a desirable and alternative plasmonic material for advanced nanoplasmonic sensing technologies.

## Keywords

Nanomaterials, plasmon sensitivity, bimetallic nanocomposites, metallic nanoalloy

## Introduction

Nanoplasmonics relentlessly exploits for light-driven applications from nano-scale biological sensing to optoelectronic technologies [1-3]. Nanoplasmonic sensors emerge as an alternative technology to overcome the limits of size miniaturization and detection resolution in conventional biological sensors. As commonly used plasmonic metals, gold and silver, features large negative real dielectric constant and small imaginary dielectric constant. Gold favors biological sensors and medical diagnostics due to biocompatibility and small ohmic loss, while silver is advantageous with intense electromagnetic field enhancement from larger extinction cross-section, along with narrow plasmon linewidth. The sensitivity (i.e. plasmon wave-length shift per local changes in refractive index) in plasmonic sensors highly depends on plasmon characteristics such as spectral linewidth, extinction cross-section, and electromagnetic-field intensity. Many researches employing gold and silver metals for nanoplasmonic sensors have been recently demonstrated, regarding the shape[4, 5], the size[6], the period[7], and hetero-metal compositions including metallic alloy[8]. Primarily, gold and silver nanoparticles have been widely explored, but exhibit low sensitivity: 70 nm/RIU for conventional AuNPs plasmon sensitivity and 161 nm/RIU for AgNPs plasmon sensitivity, respectively[9]. However, nanoplasmonic sensors still endeavor to improve low sensitivity by advanced nanoparticle designs and fabrication methods [6, 10].

Understanding a plasmonic metal as a prerequisite with intrinsic optical properties is crucial. Most metals are crystalline and enclose internal boundaries, known as a grain boundary[11]. When either a metal or a multi-metallic alloy is fabricated, the atoms within each growing grain are lined up, in a certain pattern. The limited number of atoms in metal nanoclusters results in discrete energy levels, allowing interactions with light in energy levels [12]. Each grain size in metallic nanocomposites or nanoalloys substantially affects physical properties such as spectral responses (*e.g.* Full Width at Half Maximum (FWHM)) of plasmon resonance. For instance, a single plasmonic metal nanoparticle or nanostructure is made by either chemical-synthesis[13] or thin film evaporation deposition[14]. A single metallic nanostructure is composed of ordered grains, called a *mono-metallic nanoensemble*. Localized surface plasmon resonance (LSPR) properties from a *metallic nanoensemble* is equivalent to a single metal nanoparticle, associated with collective plasmon resonance. In addition, mixing Au and Ag metals comes out a promising plasmonic material for advanced biological detection and imaging in visible and near-infrared optical ranges[15]. In this manner, hetero-metal nanostrucutres (*e.g.* Au/Ag hetero nanodimer[16] and bimetallic nanoblocks[17]), bimetallic nanocomposites[18], and bimetallic nanoalloy[8] have been extensively proposed. Choi *et al.* reported that Au-Ag nanorod multiblocks enable the detection of dopamine using quadrupole plasmon mode. Bimetallic nanoblocks also yield multiple surface plasmon bands[17]. Nonetheless, it is equivocal to speculate the optical properties in different combinations of Au and Ag at a subwavelength nano-scale regime. In addition, the intermediate state between a mono-metallic nanoensemble and a metallic nanoalloy still remains veiled. Recently, Park *et al.* reported Au/Ag bimetallic nanocomposites fabricated on a paper substrate and showed diverse feasibilities of biological applications including Metal-Enhanced Fluorescence (MEF) and Surface-Enhanced Raman Spectroscopy (SERS)[18]. A metallic nanoalloy is fabricated by thermal annealing process, pursuing the new permittivity of alloy materials to control a desirable plasmon resonance wavelength. On the contrary, bimetallic nanocomposites have a quite small grain size, ranging from sub-nanoscale to Angstrom [12, 14]. The spectral responses of bimetallic nanocomposites depending on the grain or the unit size are still unknown as well. Here we report Au/Ag bimetallic nanocomposites as a substantially high sensitive material, introducing three-dimen-sional unit approach for various plasmonic materials of *mono-metallic nanoensembles*, *bimetallic nanocomposites*, and *bimetallic nanoalloys* composed of Au and Ag. In addition, spectral responses of various plasmonc materials with Au and Ag are analytically examined to successfully achieve plasmonic sensing capability in the visible regime.

## Results and Discussion

**Spectral responses of Au mono-metallic nanoensembles and Au/Ag bimetallic nanocomposites using a unit approach**

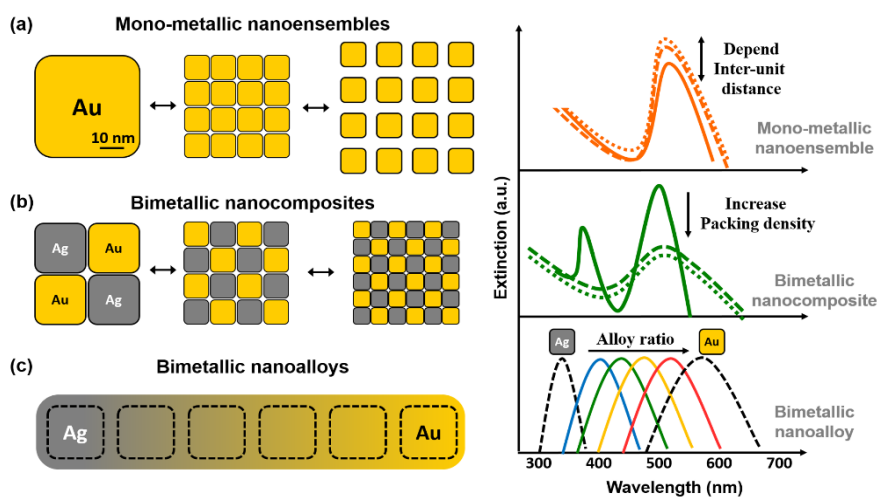

**Fig. 1.** Exemplary extinction spectra of mono-metallic nanoensembles, Au/Ag bimetallic nanocomposites, and bimetallic nanoalloys, depending on the inter-unit distance, packing density (i.e. unit size), and alloy ratio, along with cross-sectional schematics of equivalent plasmonic materials.

Plasmonic materials categorize *monometallic nanoensembles*, *bimetallic nanocomposites*, and *bimetallic nanoalloys*. Figure 1 clearly describes exemplary outcomes of spectral properties of each plasmonic material, resulting from numerical modeling using a unit approach. One of monometallic nanoensembles, Au nanoensembles materially arrange a bulk nanometal, Au nanoensemble composed of tightly-packed nanounits spaced with 1 nm, and Au nanoensemble composed of equivalent nanounits spaced with 5 nm, respectively in Fig. 1(a). A mono-metallic nanoensemble exhibits intrinsic plasmon resonance peak resulting from material property of a given metal, gold, regardless of unit packing condition. Spectral linewidth of plasmonic ensemble clusters is given by a convolution of individual nanoparticle linewidth[19]. A single metallic nanoparticle plasmon linewidth is strongly associated with damping, which follows the interband and intraband excitation of electron-hole pairs, depending on metal materials[20]. Like a single metallic nanoparticle, Au nanoensembles packed by sub-nanometer units consistently show almost the same plasmon resonance wavelength, independent of inter-unit distance. On the other hand, bimetallic nanocomposites functionalize exceptional optical responses, as illustrated in Fig. 1(b). Au/Ag bimetallic nanocomposites exclusively feature distinct plasmon resonance wavelength, which is different from bimetallic nanoalloys in Fig. 1(c). Extinction spectra in bimetallic nanocomposites mainly depend on the unit size and packing density, gradually converging multiple extinction peaks to a single extinction peak. These results are numerically demonstrated by introducing a unit approach in 3D FDTD modeling.

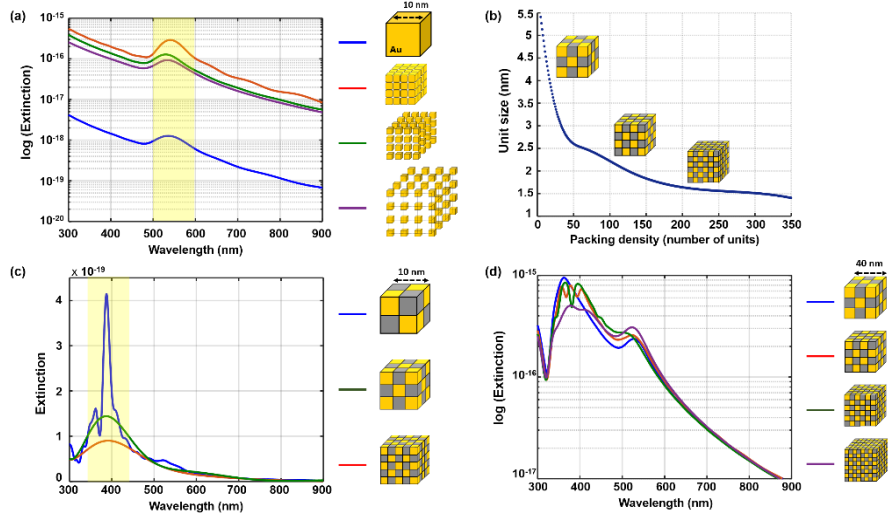

**Fig. 2.** Calculated extinction spectra of Au nanoensembles and Au/Ag bimetallic nanocomposites, depending on the packing condition (inter-unit distance, packing density, and unit size). (a) a 10 nm bulky Au nanocube (blue line) and corresponding Au nanoensembles (packing density: 64 units) depending on the inter-unit distances (0 nm, 1 nm, and 5 nm) (b) dependency of a unit size to a packing density (c) 10 nm Au/Ag bimetallic nanocomposites depending on the packing density: 8 (2x2x2) units, 27 (3x3x3) units, and 125 (5x5x5) units (d) 40 nm Au/Ag bimetallic nanocomposites depending on the packing density, 27 (3x3x3) units, 64 (4x4x4) units, 125 (5x5x5) units, and 512 (8x8x8) units.

Localized mesh technique (*i.e.* localized mesh size is 0.1 x 0.1 x 0.1 nm in $x$, $y$, and $z$ dimensions) is used to validate a unit approach in following calculations. Figure 2 shows calculated extinction spectra of Au nanoensembles and Au/Ag bimetallic nanocomposites, depending on the packing condition and unit size. Au nanoensembles identify four different types with a 10 nm cubic dimension: a Au nanocube, a tightly packed Au nanoensemble with packing density (64 units which corresponds to a 2.5 nm unit size), a 1 nm spaced Au nanoensemble with identical 64 units and 2.5 nm unit size, and a 5 nm spaced Au nanoensemble with identical 64 units and 2.5 nm unit size. In Fig. 2(a), identical extinction spectra featuring the plasmon resonance wavelength near 520 nm in wavelength is coherently observed, independent of a single Au bulky nanocube, and variously packed Au nanoensembles as a collective plasmon resonance[21]. The magnitude of extinction results from the unit size and inter-coupled capacitive electric field intensity between ensemble units [19]. Unlike mono-metallic nanoensembles, bimetallic nanocomposites strongly depend on the packing density as well as the unit size, which has inter-correlation as shown in Fig. 2(b). In this assumption, Au/Ag bimetallic nanocomposites whose total dimension is 10 x 10 x 10 nm volume cube, show consistently a single distinct plasmon resonance wavelength in Fig. 2(c). However, the spectral linewidth of Au/Ag bimetallic nanocomposites strongly depends on the packing density and unit size: as the packing density increases (i.e. the unit size decreases from 10 nm to 2.5 nm), plasmon linewidth broadening appears and dual plasmon resonance peaks converge to single plasmon resonance peak. Converged single plasmon resonance wavelength of Au/Ag bimetallic nanocomposites is closer to the localized plasmon wavelength of a single Ag nanoparticle, due to larger extinction cross-section of silver material. In addition, the maximum effective unit size is 2.5 nm in Au/Ag bimetallic nanocomposites within a subwavelength range, emphasizing a converged single plasmon wavelength. According to numerical results, the unit size of 2.5 nm is enough to get collective plasmon resonance in Au/Ag bimetallic nanocomposites, with the mesh size of 0.1 nm. Likewise, the plasmon wavelength of a bigger Au/Ag bimetallic nanocomposite, which has the dimension of 40 x 40 x 40 nm volume cube, is a little red-shifted from a 10 nm Au/Ag bimetallic nanocomposite, but almost similar spectral window, ranging from 380 nm to 410 nm in Fig. 2(d). However, the extinction spectra is more fluctuated, indicating a broad plasmon linewidth. It implies that broadening a plasmon linewidth in larger Au/Ag bimetallic nanocomposites is strongly associated with the unit size and total nanocomposite dimension, managing the degree of scattering[22]. According to whole extinction spectra of 10 nm Au/Ag bimetallic nanocomposites and 40 nm Au/Ag bimetallic nanocomposites, Au/Ag bimetallic nanocomposites enable Ag-dominant collective plasmon resonance. In analogy with bimetallic nanoalloy, the composition ratio of Au/Ag bimetallic nanocomposites functionalize a desirable plasmon resonance spectral regime in Fig. 3. To clearly identify each unit, a 40 nm nanocube coarsely packed by 27 units is composed of Au and Ag units, from a Au-dominant nanocomposite to a Ag-dominant nanocomposite in Fig. 3(a). The cross-sectional electric field distributions ($z$ component) of each Au/Ag composition ratio distinctly shows that Ag units give rise to stronger field intensity than Au units in Fig. 3(b). However, interactions between Au and Ag units yield hybrid symmetric dipolar mode. In Fig. 3(c), extinction spectra depending on the composition ratio in Au/Ag bimetallic nanocomposites follow the primary extinction resonance peak of silver and gold, respectively. Accordingly, composition ratio and arrangement of Au and Ag units in Au/Ag bimetallic nanocomposites can engineer spectral responses analogous with bimetallic nanoalloys.

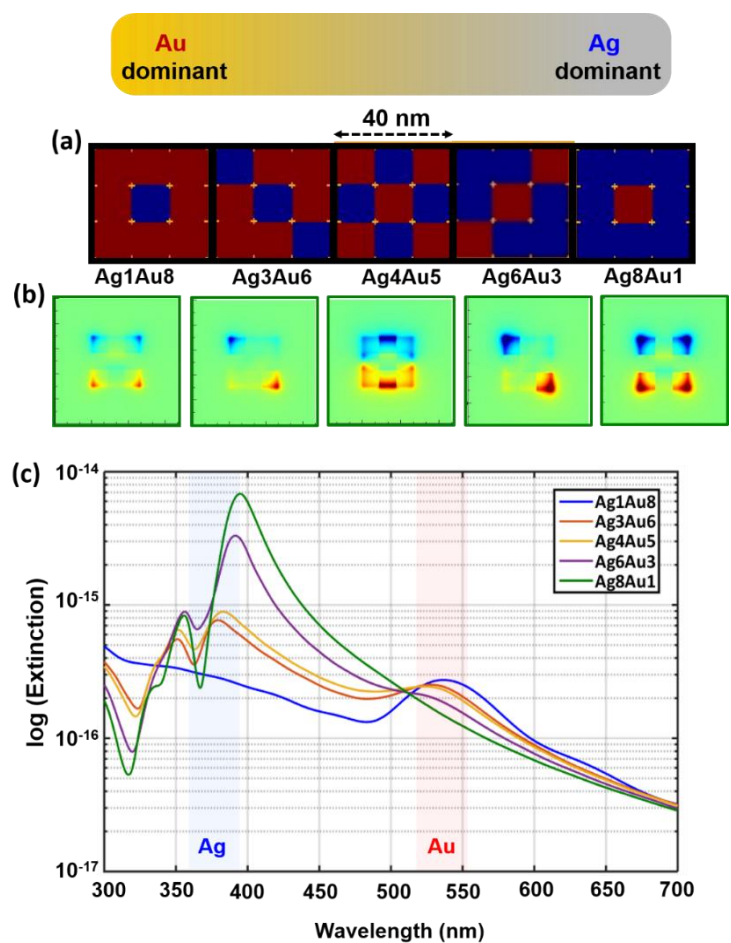

**Fig. 3**. Composition ratio of Au/Ag bimetallic nanocomposites coarsely packed by 27 units (3x3x3) (a) material index distributions from Au dominant to Ag dominant bimetallic nanocomposites (b) corresponding electric field distribution (z-component) and (c) extinction spectra with respect to the composition ratio of Au/Ag bimetallic nanocomposites.

**Improved plasmon sensitivity of Au/Ag bimetallic nanocomposites: A highly sensitive plasmonic material**

Conventional nanoplasmonic sensors have continuously attempted size miniaturization and sensitivity improvement for biological applications. Bimetallic nanostructures overcome low sensitivity and biocompatibility, taking advantage of

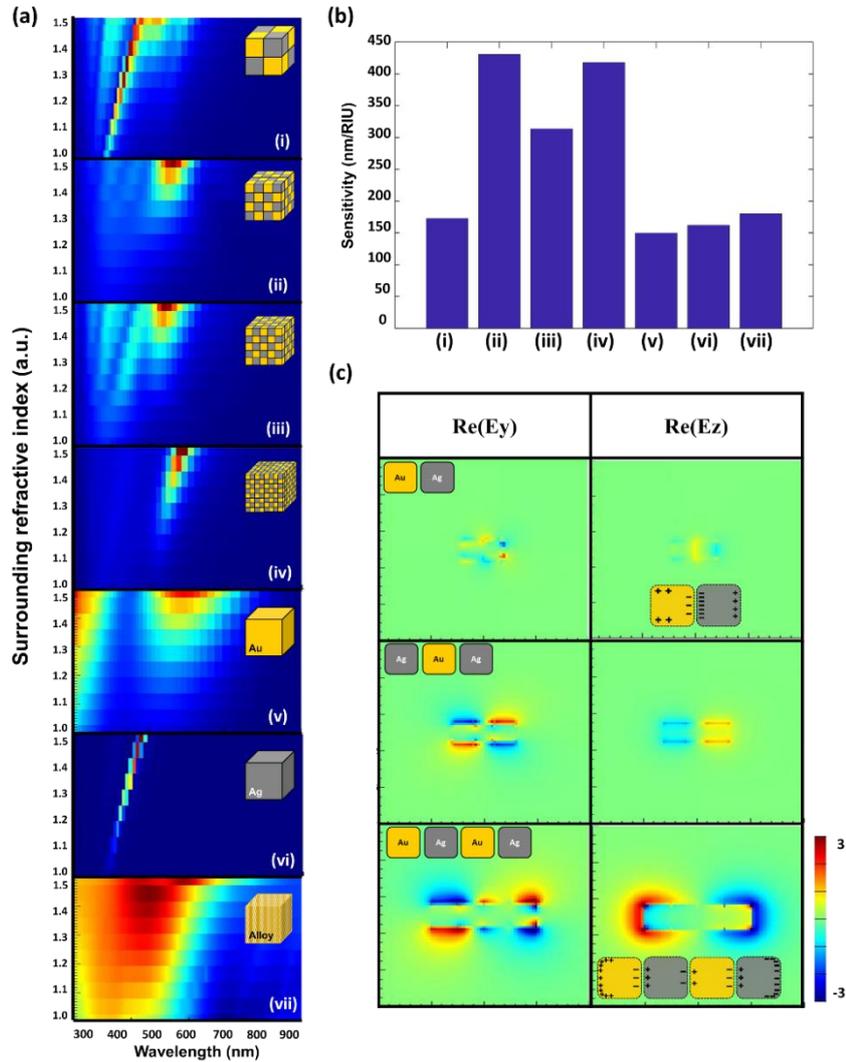

**Fig. 4.** Refractive index sensing performances: (a) Calculated spectra depending on the surrounding refractive index, which provide corresponding sensitivity which is defined as the wavelength shift divided by the surrounding refractive index variation. Each plasmonic material is 10 nm Au/Ag bimetallic nanocomposites with packing density of (i) 8, (ii) 64, (iii) 125, (iv) a 40x40x40 nm Au/Ag bimetallic nanocomposite with the packing density of 512, (v) a 40 nm Au nanocube, (vi) Ag nanocube, and (vii) a 40 nm Au/Ag bimetallic nanoalloy cube. (b) Comparison of sensitivity in each material (c) Coupled electric field distributions of *y* and *z* component depending on the coupled Au/Ag bimetallic nanounits in a row and surface charge redistribution models of asymmetric mode (a Au-Ag nanounit) and extensive symmetric mode (a Au-Ag-Au-Ag nanounit).

various plasmonic metal composition [23, 24]. In Fig. 4(b), refractive-index sensitivity of Au/Ag bimetallic nanocomposites, depending on the packing density and whole size, numerically compare with equivalent Au- and Ag- nanocube and Au/Ag bimetallic nanoalloy, respectively. From the (i)-(vii) in Fig. 4(a), the inverse slope of calculated spectra depending on the surrounding refractive index is also referred to refractive index sensitivity. In comparison, Au/Ag bimetallic nanoalloy shows slightly higher sensitivity (180 nm/RIU, (vii) in Fig. 4(a)) than a Au nanocube (150 nm/RIU, (v) in Fig. 4(a)) and a Ag nanocube (162 nm/RIU, (vi) in Fig. 4(a)). Especially, the 10 nm Au/Ag bimetallic nanocomposite packed by 64 units and 40 nm Au/Ag bimetallic nanocomposite packed by 512 units in (ii) and (iv) of Figure 4, respectively, show substantially higher sensitivity, approximately 3 fold of equivalent Au and Ag nanocubes ((ii): 430 nm//RIU, (iv): 418 nm/RIU). Quantitative improvement of refractive index sensitivity maximizes up to 186% using a single Au/Ag bimetallic nanocomposite. In addition, FWHM indicates the degree of sensing accuracy to verify spectroscopic sensors. A 40 nm Au/Ag bimetallic nanocomposite packed by 512 units in Fig. 4(d) provides much narrower FWHM as well as excellent FOM in a given comparison (see Table 1). On the other hand, the the 10

| Material | FWHM (nm) | FOM |
|---|---|---|
| (i) | 200 | 0.86 |
| (ii) | 168.8 | 2.55 |
| (iii) | 178.1 | 1.76 |
| (iv) | 56.3 | 7.43 |
| (v) | 312.5 | 0.48 |
| (vi) | 93.8 | 1.72 |
| (vii) | 302 | 0.6 |

(FOM is defined as the ratio of plasmon sensitivity to FWHM, *FOM = Sensitivity (nm/RIU) / FWHM (nm)*).

**Table 1. Plasmonic sensing capabilities of various plasmonic materials composed of Au and Ag**

nm Au/Ag bimetallic nanocomposite packed by 125 units in Fig. 4(iii) has lower sensitivity than (ii) and (iv). To reveal why a certain Au/Ag bimetallic nanocomposite provides high sensitivity, Au/Ag bimetallic nanocomposites numerically analyze coupled electric-field distributions ($y$ and $z$ components) per units in Au/Ag bimetallic nanocomposites in Fig. 4(c). Bimetallic nanocomposites are composed of sub-nanoscale metallic nanounits. In general, dipolar resonance in mono-metal nanounits is dominant[25]. Unlike mono-metallic coupled nanounits, a Au/Ag bimetallic coupled nanounit exhibits abnormal field distribution, due to unbalanced metallic properties in Fig. 4(c). As the number of bimetallic nanounits increase from a Au-Ag nanounit to a Au-Ag-Au-Ag nanounit, induced surface charges are redistributed in each metallic nanounit from an asymmetric mode to an expanded symmetric mode. In a Au-Ag-Au-Ag nanounit, symmetrically dipolar resonance mode is observed. This result can establish Au/Ag bimetallic nanocomposites composed of Au-Ag-Au-Ag quadrupole nanounits, which corresponds to in Fig.4(a) (ii) and (iv), featuring substantially a higher sensitivity than conventional plasmonic materials due to enhanced plasmon resonance intensity.

## Conclusions

We report that Au/Ag bimetallic nanocomposites offer distinct plasmonic properties beyond Au, Ag, and Au/Ag metallic nanoalloy. Using a unit approach, Au nanoensembles and Au/Ag bimetallic nanocomposites are numerically explored, depending on the packing density and composition ratio. An effective unit size of bimetallic nanocomposites in a subwavelength nanostructure is 2.5 nm tiny enough to feature clearly bimetallic nanocomposites, converging double plasmon resonance wavelength peaks from gold and silver into a singular collective plasmon resonance wavelength. As a result, increasing packing density (i.e. decreasing a unit size) enables stable plasmonic responses resulting from collective plasmon resonances in both Au nanoensembles and Au/Ag bimetallic nanocomposites. Unlike Au/Ag bimetallic nanoalloys, plasmon wavelength of Au/Ag bimetallic nanocomposites is proximal to the Ag plasmon wavelength, due to larger Ag extinction. Besides, Au/Ag bimetallic nanocomposites featuring quadrupole nanounits are capable of exceptional sensitivity (approximately 3 fold of conventional plasmon sensitivity) and FOM (4.3 fold of conventional plasmon FOM), compared to conventional plasmonic metals and Au/Ag bimetallic nanoalloys. This study of Au/Ag bimetallic nanocomposites along with mono-metallic nanoensembles can offer a deep comprehension of plasmonic materials and give a direct guideline for non-invasive and high-resolution plasmonic sensing technologies and photonic applications.

## Funding

This work was supported by Samsung Research Funding Center of Samsung Electronics under Project Number SRFC-IT1402-51 and National Research Foundation of Korea (NRF) Ministry of Science, ICT & Future Planning under Project Number 2018029899.